\documentclass[aps,prl,twocolumn,showpacs,superscriptaddress]{revtex4-1}
\usepackage{graphicx}

\begin{document}

\title{Disordered dimer state in electron-doped Sr$_{3}$Ir$_{2}$O$_{7}$}

\author{Tom Hogan}
\affiliation{Department of Physics, Boston College, Chestnut Hill, Massachusetts 02467, USA}
\affiliation{Materials Department, University of California, Santa Barbara, California 93106, USA.}
\author{Rebecca Dally}
\affiliation{Department of Physics, Boston College, Chestnut Hill, Massachusetts 02467, USA}
\affiliation{Materials Department, University of California, Santa Barbara, California 93106, USA.}
\author{Mary Upton}
\affiliation{Advanced Photon Source, Argonne National Laboratory, Chicago, Illinois, 60439 USA}
\author{J. P. Clancy}
\affiliation{Department of Physics, University of Toronto, Toronto, Ontario, Canada M5S 1A7}
\author{Kenneth Finkelstein}
\affiliation{Cornell High Energy Synchrotron Source, Cornell University, Ithaca, New York 14853}
\author{Young-June Kim}
\affiliation{Department of Physics, University of Toronto, Toronto, Ontario, Canada M5S 1A7}
\author{M. J. Graf}
\affiliation{Department of Physics, Boston College, Chestnut Hill, Massachusetts 02467, USA}
\author{Stephen D. Wilson}
\email{stephendwilson@engineering.ucsb.edu}
\affiliation{Materials Department, University of California, Santa Barbara, California 93106, USA.}

\begin{abstract}
Spin excitations are explored in the electron-doped spin-orbit Mott insulator (Sr$_{1-x}$La$_{x}$)$_3$Ir$_2$O$_7$.  As this bilayer square lattice system is doped into the metallic regime, long-range antiferromagnetism vanishes, yet a spectrum of gapped spin excitation remains.  Excitation lifetimes are strongly damped with increasing carrier concentration, and the energy integrated spectral weight becomes nearly momentum independent as static spin order is suppressed.  Local magnetic moments, absent in the parent system, grow in metallic samples and approach values consistent with one $J=\frac{1}{2}$ impurity per electron doped.  Our combined data suggest that the magnetic spectra of metallic (Sr$_{1-x}$La$_{x}$)$_3$Ir$_2$O$_7$ are best described by excitations out of a disordered dimer state.  
\end{abstract}


\maketitle
Models of Heisenberg antiferromagnets on a bilayer square lattice have generated sustained theoretical and experimental interest due to their rich variety of ground states \cite{PhysRevLett.72.2777, doi:10.1143/JPSJ.65.594, PhysRevLett.78.3019, PhysRevLett.80.5790, PhysRevB.51.16483}.  In zero field, an instability occurs above a critical ratio of interlayer to intralayer magnetic exchange that transitions spins from conventional antiferromagnetism into a dimer state comprised of spin singlets \cite{PhysRevLett.72.2777,PhysRevB.61.3475, PhysRevB.61.3475}.  These singlets may interact and form the basis for numerous uncoventional ground states such as valence bond solids \cite{PhysRevB.69.224416,PhysRevB.85.180411}, quantum spin liquids \cite{PhysRevLett.80.5790}, Bose-glass \cite{PhysRevLett.95.207206}, and other quantum disordered states \cite{PhysRevB.85.180411}.  Realizations of bilayer systems inherently near the critical ratio of interlayer to intralayer coupling however are rare, primarily due to orbital/exchange anisotropies strongly favoring either interplane or intraplane exchange pathways in accessible compounds \cite{PhysRevLett.87.217201,PhysRevB.54.16172,PhysRevB.55.8357}.   

$J_{\mathrm{eff}}=\frac{1}{2}$ moments are arranged onto a bilayer square lattice within the $n=2$ member of the Sr$_{n+1}$Ir$_{n}$O$_{3n+1}$ Ruddlesden-Popper series, Sr$_3$Ir$_2$O$_7$ (Sr-327) \cite{SUBRAMANIAN1994645}. The strong spin-orbit coupling inherent to the Ir$^{4+}$ cations in cubic ligand fields renders a largely three dimensional spin-orbit entangled wave function \cite{PhysRevLett.102.017205, PhysRevLett.112.026403}.  This combined with the extended nature of its $5d$ valence electrons presents Sr-327 as an interesting manifestation of the bilayer square lattice---one where appreciable interlayer coupling potentially coexists with strong intralayer exchange inherent to the single layer analogue Sr$_2$IrO$_4$ (Sr-214) \cite{PhysRevLett.108.177003}.

While its ground state is antiferromagnetic (AF) \cite{0953-8984-24-31-312202,PhysRevLett.109.037204,PhysRevB.86.100401}, measurements of magnetic excitations in Sr$_3$Ir$_2$O$_7$ observe anomalous spectra with large spin gaps ($\Delta E\approx 90$ meV) whose values exceed that of the single magnon bandwidth \cite{Kim, Sala}.  This has led to models shown in Figs. 1 (a) and (b) that cast the underlying exchange into two extremes:  A linear spin wave approach (LSW) with a large anisotropy gap and predominantly intraplane exchange \cite{Kim} versus a bond operator (BO) mean field approach \cite{PhysRevB.59.111} with a dominant interplane, dimer-like, exchange \cite{2012arXiv1210.1974M, Sala}.  Recently, an additional excitation attributed to a longitudinal mode associated with triplon excitations was observed supporting the latter approach\cite{PhysRevB.52.3521}.       

The comparable ability of both LSW and BO approaches to capture major features of the magnetic spectra of Sr-327 invites further study.  In particular, considerable insight can be gained by probing the evolution of spin dynamics as static AF order is suppressed.  Recent work has shown that, unlike Sr-214, Sr-327 can be driven into a homogenous metallic state with no static spin order via La-substitution \cite{Hogan,Xiang}.  Local moment behavior, notably absent in parent Sr-327 \cite{0953-8984-19-13-136214}, appears in these electron-doped samples and hints at an unconventional metallic state \cite{Hogan}.      
 
Here we utilize resonant inelastic x-ray scattering (RIXS) to explore the spin dynamics of (Sr$_{1-x}$La$_{x}$)$_3$Ir$_2$O$_7$ as it transitions from an AF insulator into a paramagnetic metal.  Beyond $x=0.04$ ($6\%$ electrons/Ir), AF order vanishes, yet robust magnetic excitations persist deep into the metallic regime.  Excitations become overdamped as carriers are introduced, yet the large spin gap inherent to the AF parent state survives into the disordered regime. The spectral weight of magnons in the metallic state becomes nearly momentum independent and exhibits a dispersion best described using the BO representation appropriate for a dimer state \cite{Sala}.  Supporting this, static spin susceptibility measurements resolve the emergence of local moments which grow with increasing La-content and are consistent with a picture where each electron doped breaks a dimer and creates an uncompensated moment.   Our aggregate data are best understood in the framework of a disordered dimer state emergent upon electron substitution in La-doped Sr-327.       

\begin{figure}
\includegraphics[scale=.25]{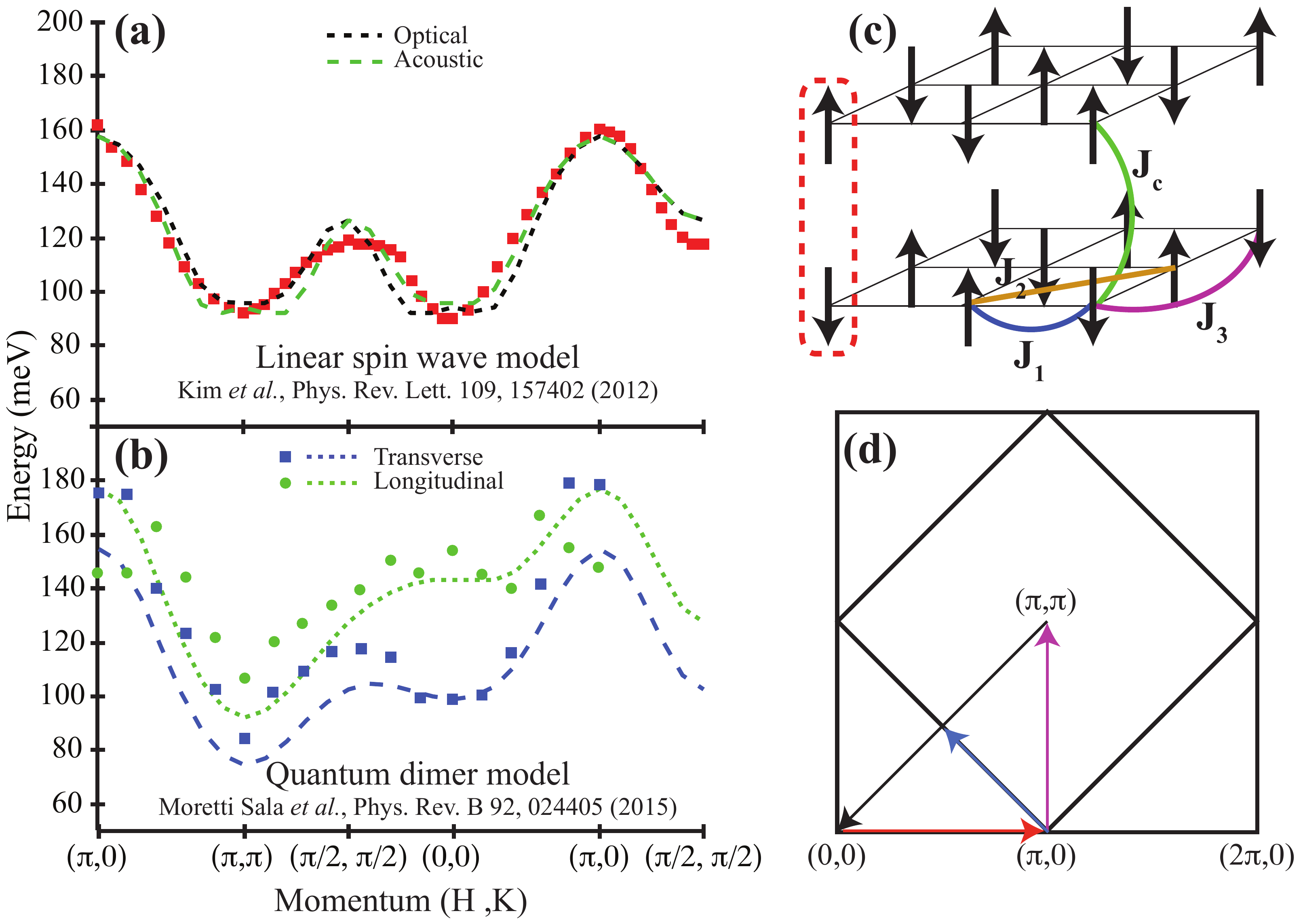}
\caption{Magnon dispersion relations for Sr-327 plotted for (a) LSW model of Kim \textit{et al.}, with data reproduced at $L=26.5$ r.l.u. from Ref. \cite{Kim} and (b) the BO model of Moretti Sala \textit{et al.}, with data reproduced at $L=28.5$ r.l.u. from Ref. \cite{Sala}.  (c)  Exchange terms within the Heisenberg models of the square lattice bilayer with dimer outlined by dashed red line (d)  Illustration of momentum cuts in panels (a) and (b) across the AF zone}
\end{figure}

Resonant inelastic x-ray scattering (RIXS) measurements were performed at 27-ID-B at the Advanced Photon Source at Argonne National Lab and resonant elastic x-ray scattering (REXS) measurements were performed at C1 at CHESS.  Details regarding the experimental setups are in supplementary information \cite{Supplemental}.  Inelastic spectra were collected near the Ir L$_3$ edge (11.215 keV) in a geometry with an energy resolution at the elastic line $\Delta E_{res}=32$ meV \cite{Supplemental}.  RIXS spectra were collected at $T=40$ K for two La-doped Sr-327 concentrations: $x=0.02$, an AF insulator ($T_{AF}\approx240$ K) and $x=0.07$, a paramagnetic metal \cite{Hogan}.  Momentum positions are denoted using the in-plane $(H,K)$ wave vectors in the approximate tetragonal unit cell ($a\approx b\approx 3.90 \AA$) and, unless stated otherwise, momentum scans were collected at $L=26.5$ [r.l.u.].    

Representative spectra for both $x=0.02$ and $x=0.07$ samples are shown in Fig. 2.  Zone center \textbf{Q}$=(\pi, \pi)$ and zone boundary \textbf{Q}$=(\pi, 0)$ cuts are shown with the elastic line ($E$), single magnon ($M$), proposed multimagnon ($M^*$), and $d-d$ excitations ($D$) shaded.  Excitations were fit to a Lorenztian of the form $I_Q(E)=\frac{2A}{\pi}\frac{\Gamma_Q}{4(E-E_Q)^2+\Gamma_Q^2}$ multiplied by the Bose population factor $(1-e^{\frac{-E}{k_B T}})$. The inverse lifetime values $\Gamma_Q$ for all excitations were substantially greater than the instrumental resolution.        
 
\begin{figure}
\includegraphics[scale=.45]{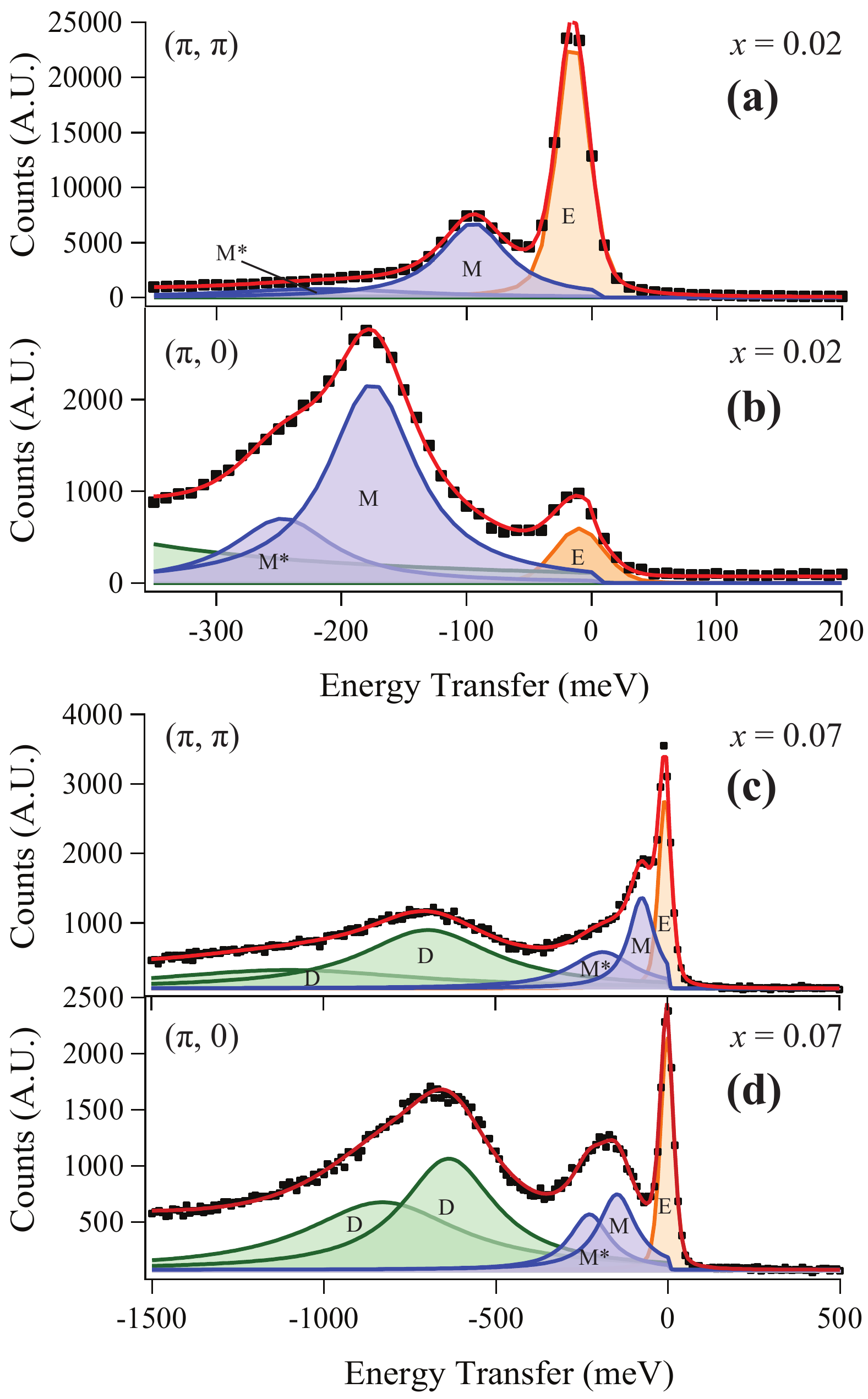}
\caption{Representative energy scans collected at $40$ K at fixed \textbf{Q} for samples with $x=0.02$ and $x=0.07$.  Panels (a) and (b) show scans performed at the AF zone center ($\pi, \pi$) and zone boundary ($\pi$, 0) for AF insulating $x=0.02$ respectively while panels (c) and (d) show the same scans for paramagnetic, metallic $x=0.07$.  Features labeled $E$, $M$, $M^{*}$, and $D$ denote scattering from the elastic line, single magnon, multimagnon, and $d-d$ excitations respectively.}
\end{figure}

The dispersions of the $M$ and $M^*$ peaks along the high symmetry directions illustrated in Fig. 1 (d) are plotted for both samples in Fig. 3.  Energies of $M$ (squares) and $M^*$ (circles) peaks are shown with the $\Gamma_Q$ associated with $M$ peaks illustrated via the larger shaded regions.  Only one feature associated with a single magnon excitation could be identified, and no additional acoustic/optical branches associated with spin waves from a bilayer or longitudinal modes associated with triplon excitations were isolated. Similar to the parent material, the $M$ peaks were independent of $L$ within the zones explored \cite{Kim,Supplemental}. Any weak additional modes are obscured due to the overdamping of the $M$ excitations as carriers are introduced \cite{Supplemental}---an effect which partially convolves the $M$ and $M^*$ features. 

Despite the absence of this second mode, tests can still be made within the RIXS spectra regarding the suitability of the LSW and BO approaches to the bilayer square lattice Heisenberg Hamiltonian with inplane exchange constants $J_1, J_2, J_3$, interplane exchange $J_c$, and an anisotropy term $\theta$ as illustrated in Fig. 1 (c) \cite{Sala, Kim}.  Specifically, the data show that the gap energies of $M$-peaks at the ($\pi, \pi$) and ($0, 0$) positions become increasingly inequivalent upon doping.  For the $x=0.07$ sample, the AF zone center ($\pi, \pi$) gap value decreases to $E_{\pi , \pi}=73\pm4$ meV whereas the $\Gamma$-point ($0, 0$) gap remains nearly unchanged from the parent system at $E_{0,0}=89\pm4$ meV \cite{Supplemental}.  The differing energies of the $M$ peaks at these two points suggest that a simple LSW model cannot account for the dispersion \cite{Kim}.  In a naive LSW approach, the combined optical plus acoustic spectral weight should remain degenerate at the ($\pi, \pi$) and $(0, 0)$ positions, which for the $x=0.07$ sample would violate the assumption that both an acoustic and optical mode are convolved within the largely $L$-independent $M$ excitations \cite{Supplemental, Kim}.  The BO approach however allows for nondegenerate spectral weight at these positions through inequivalent transverse mode $E_{\pi, \pi}$ and $E_{0, 0}$ gap values whose ratio is governed by the anisotropy term $cot(\theta)=E_{0,0}/E_{\pi,\pi}$.  Therefore, to parameterize the dispersion in electron-doped Sr-327 samples, the BO model was utilized \cite{2012arXiv1210.1974M,Sala}.        
  
\begin{figure}
\includegraphics[scale=.3]{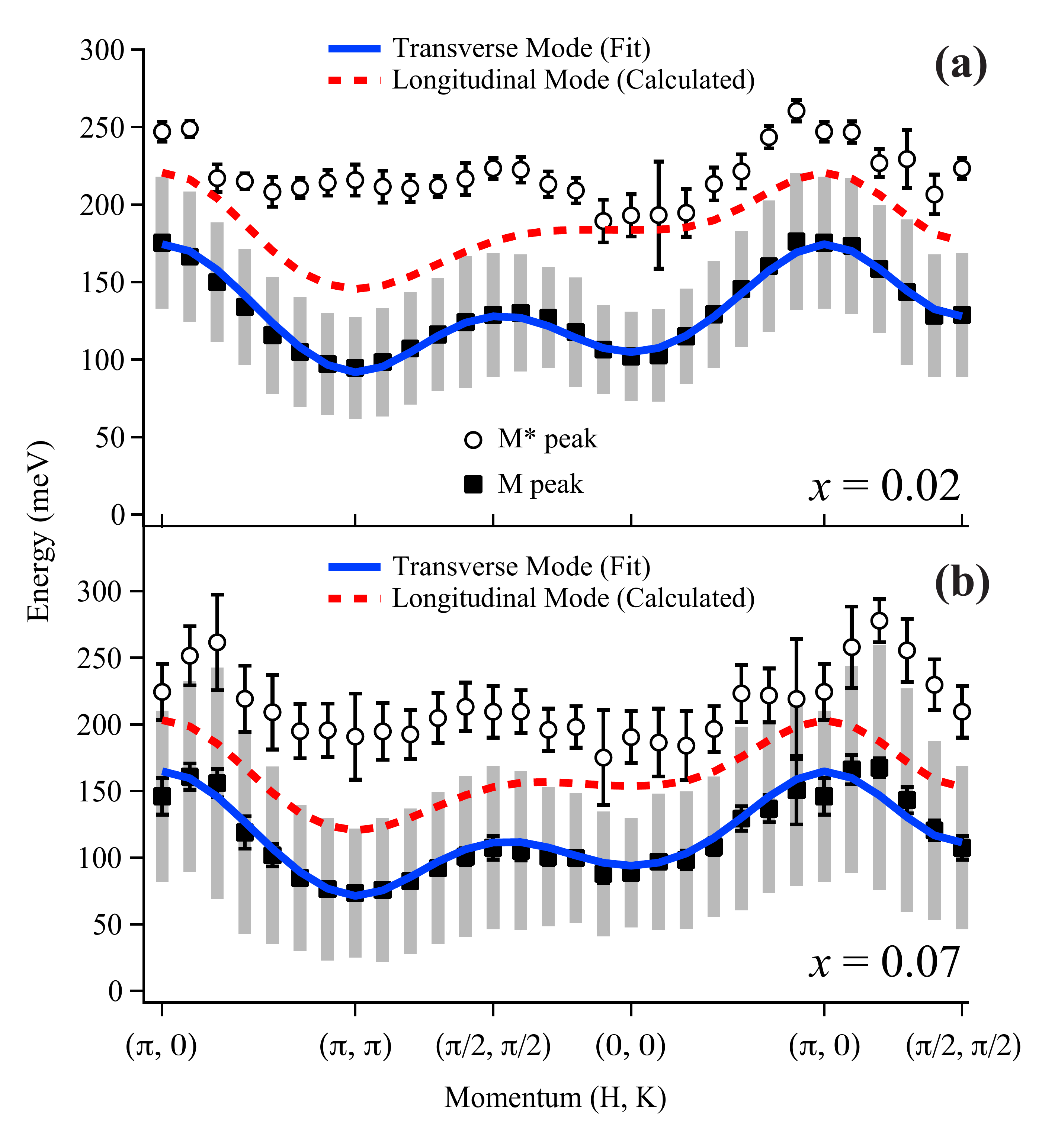}
\caption{Dispersion of $M$ and $M^{*}$ features for (a) $x=0.02$ and (b) $x=0.07$ samples.  $M$ and $M^{*}$ peaks are plotted as squares and circles respectively, each with accompanying errors.  The larger shaded regions about the $M$-dispersion are the excitations' FWHM.   Solid lines denote fits to the transverse modes using the BO model and dashed lines denote the predicted positions of longitudinal modes.}
\end{figure}

Fits using the BO generated dispersion relations along the pathways illustrated in Fig. 1 (d) are shown as solid lines in Figs. 3 (a) and (b).  Due to the suppressed spectral weight expected for the longitudinal mode \cite{Supplemental,Sala} and the broadened $\Gamma$ values inherent to doped samples, the predicted longitudinal branches lie convolved either within the FWHM of the $M$ mode or $M^*$ feature. Fits were therefore performed only to the transverse modes' dispersion, and the predictions for the accompanying longitudinal modes are plotted for reference.  Using this parameterization, the coupling constants evolve from $J_1=37.7$ meV, $J_2=-14.0$ meV, $J_3=4.8$ meV, and $J_c=87.6$ meV for the $x=0.02$ sample to $J_1=29.1$ meV, $J_2=-17.0$ meV, $J_3=5.2$ meV, and $J_c=80.1$ meV for the $x=0.07$ sample.  The anisotropy term decreased slightly from $\theta=41.2$ for $x=0.02$ to $\theta=37.2$ for $x=0.07$.  The inverse lifetimes of the $M$-excitations are largely $Q$-independent and increase from an average value of $\Gamma_{avg}=75$ meV for $x=0.02$ to $\Gamma_{avg}=124$ meV for $x=0.07$ \cite{Supplemental}.             

\begin{figure}
\includegraphics[scale=.25]{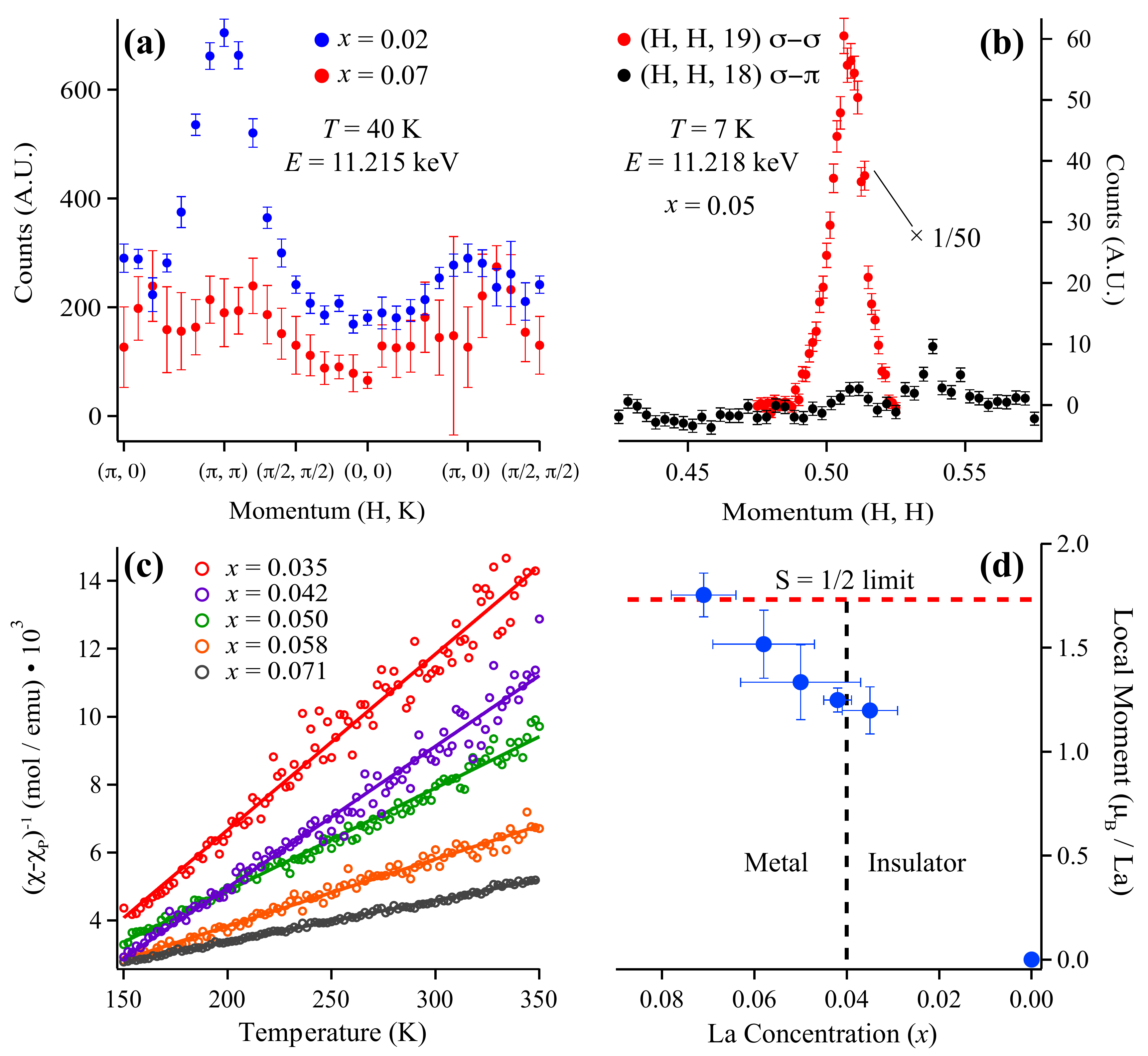}
\caption{(a) Energy integrated spectral weight of $M$ peaks across the AF zone.  Data for $x=0.02$ (blue circles) show a maximum at the zone center consistent with its AF ordered ground state.  Data for $x=0.07$ (red circles) show a nearly Q-independent response.  (b)  REXS data showing the absence of AF correlations in the $x=0.05$ sample.  Black circles denote $H$-scans through the AF position (0.5, 0.5, 18) in the $\sigma -\pi$ channel. Red circles denote the structural reflection at (0.5, 0.5, 19) in the $\sigma -\sigma$ channel scaled by $1/50$ for clarity.  Background has been removed from the data.  (c)  Curie-Weiss fits to high temperature susceptibility with a temperature independent $\chi_0$ term removed and collected under $H=5$ kOe.  (d)  Local moments $\mu_{\mathrm{eff}}$/La extracted from fits in panel (c).}
\end{figure}

While electron-doping drives a subtle shift in the $M$ dispersion, the bandwidth is largely unaffected upon transitioning from the AF insulating regime ($x=0.02$) into the paramagnetic, metallic state ($x=0.07$).  This is striking, in particular due to the reported absence of magnetic order in the $x=0.07$ sample \cite{Hogan}.  The distribution of the spectral weights of $M$ peaks in both samples further reflect this fact and are plotted in Fig. 4 (a). In the AF $x=0.02$ sample, the energy integrated weight is maximal at the magnetic zone center ($\pi$, $\pi$) as expected \cite{PhysRevB.84.020403}; however this zone center enhancement vanishes with the loss of AF order in the $x=0.07$ sample.  

In order to further search for signatures of remnant short-range order in the metallic regime, REXS measurements were collected at $7$ K on an $x=0.05$ crystal. Data collected at the Ir $L_3$ edge are plotted in Fig. 4 (b) showing $H$-scans through the magnetic $(0.5, 0.5, 18)$ and structural $(0.5, 0.5, 19)$ reflections in the $\sigma-\pi$ and $\sigma-\sigma$ channels respectively.  No peak is observed in the $\sigma-\pi$ channel at the expected magnetic wave vector; however the weak structural peak apparent in the $\sigma-\sigma$ channel at $(0.5 ,0.5, 19)$ gives a sense of the measurement sensitivity.  By performing identical measurements on an $x=0.02$ sample with a known AF moment $m_{AF}\approx0.31$ $\mu_B$, this places a bound of $m_{AF}<0.06$ $\mu_B$ for the metallic x=0.05 concentration \cite{Hogan, Supplemental}.  Scans along other high symmetry directions also failed to detect static short-range correlations.

The absence of static antiferromagnetism in samples with $x>0.04$ is consistent with earlier neutron diffraction measurements \cite{Hogan} and render it distinct from its single layer analogue, Sr-214.  In electron-doped Sr-214, short-range AF order survives to the highest doping levels explored $\approx 12\%$ electrons/Ir \cite{Xiang,2016arXiv160307547G} and can account for a magnon dispersion with slightly renormalized magnetic exchange \cite{2016arXiv160102172L}.  In contrast, electron-doping Sr-327 reveals gapped spin excitations that persist beyond the disappearance of AF order. While a slight increase in $J_c/J_1$ from 2.32 to 2.75 accompanies the disappearance of AF order and is naively consistent with predictions for the formation of a dimer state beyond a critical ratio of $J_c/J_1\approx 2.5$ \cite{PhysRevLett.72.2777,PhysRevB.61.3475}, the extended in-plane exchange and anisotropy terms used in the BO approach of Ref.\ \cite{Sala} as well as the presence of doped carriers necessarily modify this critical threshold \cite{PhysRevB.52.7708}.

Although doping complicates models of dimer excitations, it also provides a further test for a hidden dimer state in Sr-327.  In the simplest picture, adding an electron to the IrO$_2$ planes creates a nonmagnetic Ir$^{3+}$ site within a sea of $J=\frac{1}{2}$ moments.  For a ground state composed of uncorrelated dimers, this nonmagnetic site should break a dimer and leave an uncompensated $J=\frac{1}{2}$ moment behind.  Hence, doping the dimer state with electrons should seed nonmagnetic Ir$^{3+}$ sites and create an increasing fraction of weakly coupled, uncompensated spins within the sample. An order by disorder transition should eventually follow among these unfrustrated local moments in the $T=0$ limit \cite{doi:10.1143/JPSJ.65.2385, PhysRevLett.86.1086,PhysRevLett.103.047201}.     
          
Intriguingly, previous magnetization measurements reported an unusual Curie-Weiss (CW) response in electron-doped Sr-327 \cite{Hogan}.  This fact combined with the absence of CW behavior in the high temperature susceptibility of the parent system \cite{Supplemental} suggests a dopant induced local moment behavior.  To explore this further, magnetization measurements were performed on a series of Sr-327 samples with varying levels of La-doping.  The high temperature CW susceptibilities for each sample are plotted in Fig. 4 (c) and the local paramagnetic moments ($\mu_{\mathrm{eff}}$) are plotted as a function of La-concentration in Fig. 4 (d).  The $\mu_{\mathrm{eff}}$ extracted from CW fits grows with increasing doping, and the $\mu_{\mathrm{eff}}$ induced per La-dopant approaches that of uncompensated $J=\frac{1}{2}$ local moments.  The absence of static AF order combined with the growth of local moments in the presence of significant AF exchange supports the notion of an underlying disordered dimer state in metallic Sr-327.   

The nearly $Q$-independent energy-integrated spectral weight of the $M$-excitations in the metallic regime is also consistent with a dimer state where the intradimer coupling ($J_c$) approaches the excitation bandwidth. The small increase in the $J_c/J_1$ ratio as doping is increased from $x=0.02$ and $x=0.07$ samples is however not the likely driver for the dimer state's stabilization, in particular given that $J_c/J_1\approx3.5$ reported for the AF ordered parent system \cite{Sala} exceeds the ratios for both of the doped compounds. Additionally, structural changes driven by electron doping in Sr-327 are relatively small, and the nearly cubic ligand field of Sr-327 ($\Delta_d = 1.10\times10^{-4}$ \cite{PhysRevB.93.134110}) does not change appreciably with electron doping \cite{Hogan}.  Rather, a dimer state is likely stabilized by the critical threshold for dimer formation being driven downward via electron-doping similar to $t-J$ models of hole-doping in bilayer cuprates \cite{PhysRevB.52.7708,PhysRevB.60.15201,PhysRevLett.106.136402}.  
    
In summary, RIXS data reveal spin excitations in La-doped Sr-327 that persist across the AF insulator to paramagnetic metal transition.  Across the insulator-metal transition, static AF correlations vanish and extended LSW models fail to describe the surviving spin spectra with nondegenerate excitations at the two-dimensional AF zone center and $\Gamma$ points.  Rather a BO-based mean field approach, reflective of strong interplane dimer interactions, captures the observed dispersion and suggests a disordered dimer state in the metallic regime.  The presence of a hidden, disordered dimer state is supported by bulk magnetization data which reveal the emergence of anomalous local moments in electron-doped Sr-327 and are consistent with dopant-induced creation of uncompensated spins from broken dimer pairs.  Our results point toward an unconventional metallic state realized beyond the collapse of spin-orbit Mott state in Sr$_3$Ir$_2$O$_7$.     
  
\begin{acknowledgments}
S.D.W. thanks L. Balents for helpful discussions.  This work was supported in part by NSF award DMR-1505549 (S.D.W.), as well as by the MRSEC Program of the National Science Foundation under Award No. DMR-1121053 (T.H.). This research used resources of the Advanced Photon Source, a U.S. Department of Energy (DOE) Office of Science User Facility operated for the DOE Office of Science by Argonne National Laboratory under Contract No. DE-AC02-06CH11357.  CHESS is supported by the NSF and NIH/NIGMS via NSF award DMR-1332208.  Work at the University of Toronto was supported by the Natural Sciences and Engineering Research Council of Canada through the Discovery Grant and the CREATE program.  SQUID measurements were supported in part by NSF award DMR-1337567.
\end{acknowledgments}

\bibliography{BibTex}

\end{document}